\documentclass[12pt]{article}
\usepackage{amsfonts}
\usepackage{epic}
\usepackage{curves}
\usepackage{pstcol}

\newtheorem{example}{Example}

\let\Bbb=\bf
\let\wtd=\widetilde
\def\XX{{\hbox{\scriptsize${{\hbox{\tiny$\times$}}\atop{
\hbox{\tiny$\times$}}}$}}}
\newcommand{\ORD}[1]{\XX{#1}\XX}
\newcommand{\D}{\hbox{d}}

\usepackage{amssymb}

\newcommand{\newsection}{
\setcounter{equation}{0}
\section}
\def\appendix#1{
\addtocounter{section}{1} \setcounter{equation}{0}
\renewcommand{\thesection}{\Alph{section}}
\section*{Appendix \thesection\protect\indent
#1}
}
\newcommand{\cpict}[3]{
\dimen1=#1\advance\dimen1 by-\hsize\divide\dimen1 by-2
\vtop to #2{
\noindent\hskip\dimen1{\special{em:graph #3.bmp}}
\vfil}\hskip-2cm
}

\newcommand{\tr}{\,{\rm tr}\,}
\def\e{\E}

\def\be{\begin{equation}}
\def\ee{\end{equation}}
\def\bea{\begin{eqnarray}}
\def\eea{\end{eqnarray}}

\let\Bbb=\mathbb
\def\HH{{\Bbb H}}
\def\RR{{\Bbb R}}

\newcommand{\E}{\hbox{e}}

\newcommand\I{{i}}

\begin{document}

\title{{\bf Extension of geodesic algebras to continuous genus}
\vspace{.5cm}}
\author{{\bf L. Chekhov},\thanks{Steklov Mathematical Institute,
Gubkina 8, 117966, GSP-1, Moscow, Russia; Institute for Experimental and
Theoretical Physics, Moscow, Russia; Laboratoire Internationale Franco--Russie.
Email: chekhov@mi.ras.ru.}
{\bf J. E. Nelson},\thanks{Dipartimento di Fisica Teorica, Universit\`a degli Studi di Torino
and Istituto Nazionale di Fisica Nucleare, Sezione di Torino, via Pietro
Giuria 1, 10125, Torino, Italy, email: nelson@to.infn.it.} \
and {\bf T. Regge}\thanks{Dipartimento di Fisica, Politecnico di Torino,
Corso Duca degli Abruzzi, 10129 Torino, Italy.}
\date{ }}

\maketitle

\begin{abstract}
Using the Penner--Fock parameterization for Teichm\"uller spaces of
Riemann surfaces with holes, we construct the string-like free-field
representation of the Poisson and quantum algebras of geodesic functions
in the continuous-genus limit. The mapping class
group acts naturally in the obtained representation.
\end{abstract}

\leftline{{\bf Mathematics Subject Classification} (2000) 14D21, 46L87, 53C22.}
\medskip
\leftline{{\bf Keywords:} 3D gravity, moduli spaces,
geodesics, large-genus limit}

\newsection{Introduction}
The classical phase space of Einstein gravity on a 3D manifold is the
Teichm\"uller space of its boundary \cite{VV}. This Teichm\"uller space has a
canonical (Weil--Petersson) Poisson structure with the mapping class group as
its symmetry group. When quantizing this structure \cite{ChF,Ch-Fock-TMF}, the
algebra of observables in the corresponding quantum theory is the
noncommutative deformation of the $*$-algebra of functions whose Poisson
brackets become commutation relations with quantum parameter $\hbar$. The
symmetry group acts on the algebra of observables by automorphisms generated by
unitary operators of the quantum mapping class group. These observables are the
lengths of closed geodesics on a Riemann surface, and they were constructed
from the Teichm\"uller space coordinates using the graph technique of Penner
\cite{Penner} and Fock \cite{Fock1}. The induced geodesic algebras with Goldman
Poisson brackets \cite{Gold} are exactly those obtained in the first-order
formalism description of $2+1$-dimensional gravity by Nelson, Regge, and
Zertuche \cite{NR,NRZ}.

In Section 2, we briefly review the first-order formalism of 3D (i.e. 2+1)
gravity in which Goldman brackets appear dynamically; in Section 3, we review
the graph description of Teichm\"uller spaces of Riemann surfaces with holes
and introduce a set of geodesic functions, the observables in the theory. In
Section 4, we consider the specific basis of observables for which the
Nelson--Regge Poisson algebra is satisfied. In Section 5, using the
Penner--Fock coordinates, we construct the limiting expressions for geodesic
functions in the limit of infinite, or continuous genus and find that they
satisfy a simple string-like (or CFT-like)
representation in terms of free  fields. The
Poisson relations for these fields coincide with those of the string
coordinates. We also consider the action of the subclass of the mapping class
group transformations generated by Dehn twists along geodesics from those
considered by Nelson and Regge \cite{NRZ}, and show that these symmetries act
naturally within the above free-field representation. We construct the
quantization of the Poisson geodesic algebras in the limit of infinite genus in
Section 6.

\newsection{The first-order formalism in 3D gravity}
Here we briefly describe the approach to
$2+1$-dimensional gravity based on the first-order
formalism \cite{NR,NRZ}. For the standard Einstein--Hilbert action
(without matter) and cosmological constant $\Lambda$
\be
I=\frac{1}{16\pi G}\int_M \D^3x\sqrt{-g}(R-2{\Lambda}),
\label{C1}
\ee
the classical solutions are constant curvature spaces with topology
$M\sim[0,1]\times S$, where $S$ is a closed 2D surface.

In the first-order formalism, the fundamental independent variables are
a spin connection $w_\mu{}^{ab}$, and a local frame (or ``dreibein'')
$e^a_\mu$ satisfying

\noindent $\eta_{ab}e_\mu{}^a e_\nu{}^b=g_{\mu\nu}$, \
$\eta_{ab}=\hbox{diag\,}\{-++\}$.  Introducing the one-forms
$e^a=e_\mu{}^a \D x^\mu$ and $w^a=\frac{1}{2}\epsilon^{abc}w_{\mu bc} \D
x^\mu$ the action (\ref{C1}) becomes
\be
I=-2\int_M \left(e\wedge
dw+\frac{1}{2}e\wedge w\wedge w +\frac{{\Lambda}}{6}e\wedge e\wedge
e\right).
\label{C6}
\ee
Setting ${\Lambda}=-1/\ell^2<0$,
and introducing the real variables $A^{(\pm)a}=w^a\pm \frac{1}{\ell}e^a$,
we obtain the Chern--Simons (CS) theory with an $SO(2,1)\times
SO(2,1)$ gauge potential
\be
I[A^{(+)},A^{(-)}]=I_{CS}[A^{(+)}]-I_{CS}[A^{(-)}], \label{C8}
\ee
where
\be
I_{CS}[A]=\frac{k}{4\pi}\int_M\tr\left(A\wedge dA
+\frac{1}{3}A\wedge A\wedge A\right)
\label{C9}
\ee
is the standard CS action.

The two dynamical (spacelike)
components satisfy the Poisson brackets
\be
\left\{A_i^{(\pm)a}(x),A_j^{(\pm)b}(x')\right\}=\pm\frac{1}{\ell}\epsilon_{ij}
\eta^{ab}\delta^2(x-x'), \label{PPP}
\ee
while all of the $A^+$ variables
Poisson commute with all of the $A^-$ variables. The holonomies, or
geodesic functions, are
\be
G^{(\pm)}_\gamma=P\exp\left\{\int_\gamma
A^{(\pm)a}_iT_a \D x^i\right\}, \label{holonomy}
\ee
and their Poisson brackets were first found by Goldman \cite{Gold}. However,
these brackets and the skein relations can be obtained from different
considerations based on the graph description.

\newsection{Classical Teichm\"uller spaces. The graph technique}
Classically, Teichm\"uller spaces ${\cal
T}^{}_{g,s}$  of Riemann surfaces of genus
$g$ with $s$ holes ($s>0$) are spaces of complex structures on a
(possibly open) Riemann surface~$S$ modulo diffeomorphisms
homotopy equivalent to the identity. In the vicinity of a boundary
component, the complex structure is isomorphic as a complex
manifold either to an annulus (hole) or to a punctured disc
(puncture).

An oriented 2D surface can be continuously conformally
transformed to a constant curvature surface. The Poincar\'e
uniformization theorem states that any complex surface $S$ of a
constant negative curvature is a quotient of the upper half-plane
${\HH}_+$ endowed with the hyperbolic metric $ds^2=dzd\overline
z/(\Im z)^2$ with respect to (w.r.t.) \ the action of a discrete Fuchsian
subgroup~${\Delta}(S)$ of the automorphism group $PSL(2,\RR)$,
$$
S={\HH}_+/{\Delta}(S).
$$

Any hyperbolic homotopy class of closed curves $\gamma$ contains
a unique {\it closed geodesic} of the length
$l(\gamma)=\log\left|{\lambda_1}/{\lambda_2}\right|$, where
$\lambda_1$ and $\lambda_2$ are (different) real eigenvalues of the
element of $PSL(2,\RR)$ corresponding to~${\gamma}$. These
lengths are the desired physical observables.

The central point of the construction is the description of the
Teichm\"uller space ${\cal T}^{}_{g,s}(S)$ in terms of fat
graphs \cite{Fock1}, which we briefly describe.

Let a {\it fat graph} ${\Gamma}$ be a graph with a given cyclic
ordering of the edges entering each vertex. We take all vertices to be
three-valent, and coordinatize the Teichm\"uller space
by associating {\it real numbers\/} $Z_\alpha\in\RR$ to unoriented edges.
In this way we obtain
the set $\{Z_\alpha|\alpha\in E({\Gamma})\}$,
where $E({\Gamma})$ is the set of all (unoriented) edges of the
graph ${\Gamma}$. We also let~$V(\Gamma)$ denote the set of
vertices of a graph~$\Gamma$.

A fat graph that is embedded into an oriented surface inherits the
canonical fat structure from the surface orientation. Denote by
${\Gamma}(S)$ the set of all isotopy classes of three-valent fat graphs
embedded in the surface $S$ in such a way that the surface is shrinkable
to the image. For any $u \in {\Gamma}(S)$ and for any given edge of
$u$,  associate a standard morphism $u \mapsto v$, where
$v\in{\Gamma}(S)$ is obtained from $u$ by a {\em flip} of the edge
(shown in Figure 1 below). These morphisms, together with
symmetries of a graph, generate the {\it mapping class group}.

A three-valent graph ${\Gamma}$ of genus~$g$ with $s$ holes
therefore generates an isomorphism between the
set of points of ${\cal T}^{}_{g,s}(S)$ and the set
${\RR}^{\hbox{\tiny \#\,edges}}$ of edges of this graph endowed with
real numbers \cite{Fock1}.

In order to parameterize closed geodesics ({\it paths} over edges of the
graph) on the Riemann surface, associate the matrix $X_{Z_\alpha}\in
PSL(2,\RR)$ corresponding to a M\"obius transformation to each
edge~$\alpha$,
\be \label{XZ} X_{Z_\alpha}=\left(
\begin{array}{cc} 0 & -\e^{Z_\alpha/2}\\
                \e^{-Z_\alpha/2} & 0\end{array}\right)
\ee
and introduce the matrices for ``right'' and ``left'' turns
\be
\label{R} R=\left(\begin{array}{rr} 1 & 1
\\ -1 & 0\end{array}\right),
\qquad L\equiv
R^2=\left(\begin{array}{rr} 0
& 1\\ -1 & -1\end{array}\right).
\ee

Let a {\it closed path} in the graph~${\Gamma}$ be any oriented
path (no turning back), which starts and terminates
at the same edge of the graph. The product of
matrices standing on consecutive edges and turns of the path is
\be
\label{PZ}
P_{Z_1\dots Z_k}= LX_{Z_k}LX_{Z_{k-1}}RX_{Z_{k-2}}\dots
RX_{Z_2}LX_{Z_1}. \label{Path}
\ee

In this way one obtains \cite{Fock1}
a one-to-one correspondence between the set of conjugacy classes of oriented
paths $\{P_{Z_1\dots Z_k}\}$ and closed {\rm(}oriented\/{\rm)} geodesics
$\{\gamma\}$ on the surface~$S$. The length  $l(\gamma)$ of a geodesic is
determined from
\be
\label{geod}
G_\gamma\equiv 2\cosh (l(\gamma)/2)=\tr
P_{Z_1\dots Z_k}.
\ee
We shall call $G_\gamma$ a {\it geodesic function}. It should be identified with
the geodesic function of equation (\ref{holonomy}) when $\Lambda = -1$.

In the $\{Z_\alpha\}$ coordinates
a canonical Poisson structure (the Weil--Petersson
structure) $B_{{\mbox{\tiny WP}}}$ on ${\cal T}^{}_{g,s}(S)$
is
\be
\label{WP-PB} B_{{\mbox{\tiny WP}}} = \sum_{v\in V(\Gamma)}
\sum_{i=1}^{3}\frac{\partial}{\partial Z_{v_i}}\wedge
\frac{\partial}{\partial Z_{v_{i+1}}},
\ee
where~$v_i$, \ $i=1,2,3\ \hbox{\rm mod}\
3$, label the cyclically ordered edges incident on a vertex~$v$.
This structure is degenerate, and its Casimir functions are
exactly the lengths of the geodesics encircling holes.

The graph transformation that preserves the Poisson structure
(\ref{WP-PB}) under the {\it flip} operation and satisfies the
{\it pentagon identity} \cite{ChF} is shown in Figure 1 where
\be
\{A,B,C,D,Z\}\to\{A+\phi(Z), B-\phi(-Z), C+\phi(Z), D-\phi(-Z),-Z\}
\label{abc}
\ee
and classically $\phi(Z)=\log(\e^Z+1)$ \cite{Fock1}.

It was shown in \cite{ChF} that the transformation~{\rm(\ref{abc})}
preserves the traces of products over paths {\rm(\ref{geod})}, i.e. the
classical geodesic lengths are invariant under the action of the mapping
class group.

\vspace{10pt}

\setlength{\unitlength}{1mm}%
\begin{picture}(50,27)(-20,48)
\thicklines \put(28,70){\line( 1,-2){ 4}} \put(32,62){\line( 1,
0){28}} \put(60,62){\line( 1, 2){ 4}} \put(60,62){\line( 1,-2){
4}} \put(32,62){\line(-1,-2){ 4}} \thinlines
\put(18,62){\vector(-1, 0){  0}} \put(18,62){\vector( 1, 0){ 5}}
\thicklines \put(10,54){\line( 2,-1){ 8}} \put(10,70){\line(
0,-1){16}} \put(10,54){\line(-2,-1){ 8}} \put( 2,74){\line(
2,-1){ 8}} \put(10,70){\line( 2, 1){ 8}} \put(
4,74){\makebox(0,0)[lb]{$A$}} \put(16,74){\makebox(0,0)[rb]{$B$}}
\put(12,62){\makebox(0,0)[lc]{$Z$}}
\put(16,50){\makebox(0,0)[rt]{$C$}} \put(
4,50){\makebox(0,0)[lt]{$D$}} \put(30,54){\makebox(0,0)[lt]{$D -
\phi(-Z)$}} \put(80,54){\makebox(0,0)[rt]{$C+\phi(Z)$}}
\put(82,69){\makebox(0,0)[rb]{$B-\phi(-Z)$}}
\put(30,69){\makebox(0,0)[lb]{$A+\phi(Z)$}}
\put(47,64){\makebox(0,0)[cb]{$-Z$}}
\end{picture}

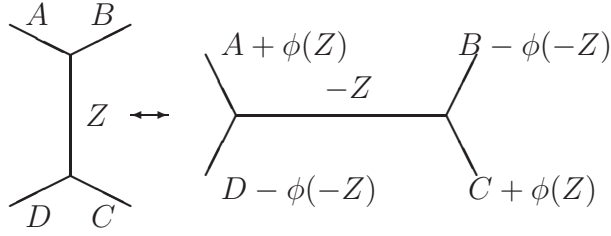
\centerline{\bf Figure 1}

The functions $G_\gamma$ (\ref{geod}) studied in \cite{Gold} satisfy a
multiplicative Poisson bracket algebra over ${\Bbb Z}$.

The classical skein relation follows from the formula
$\tr(AB)+\tr(AB^{-1})-\tr A\cdot\tr B=0$, which holds for
arbitrary $2\times 2$ matrices~$A$ and~$B$ with unit determinants. This
can be represented graphically as
\be
\setlength{\unitlength}{.6mm}%
\begin{picture}(90,20)(0,50)
\thicklines
\put(-20,48){\makebox(0,0){$\tr B$}}
\put(-20,72){\makebox(0,0){$\tr A$}}
\put(-10,50){\line(1, 1){9}} \put( 1, 61){\line(1, 1){9}}
\put(-10,70){\line(1,-1){20}}
\put(16,60){\makebox(0,0){$=$}}
\put(33,60){\makebox(0,0){$\tr AB$}}
\put(40,60){\oval(20,20)[r]} \put(63,60){\oval(20,20)[l]}
\put(64,60){\makebox(0,0){$+$}}
\put(78,60){\makebox(0,0){$\tr AB^{-1}$}}
\put(100,48){\oval(20,20)[t]} \put(100,72){\oval(20,20)[b]}
\put(110,60){\makebox(0,0){$.$}}
\end{picture}
\label{skeinclass} \ee

Using this relation we can disentangle any set of
crossing geodesics; the basis elements are therefore {\it geodesic
laminations\/} (GLs), namely sets of nonintersecting and
nonselfintersecting geodesics on~$S$.

Consider now the Poisson structure of geodesic functions. Two
nonintersecting geodesics have a trivial bracket. Because a Poisson
bracket $\{G_1,G_2\}$ of two geodesic functions satisfies the
Leibnitz rule, it suffices to consider only a single
intersection of two geodesics. Graphically, we obtain the relation
\be
\setlength{\unitlength}{.6mm}%
\begin{picture}(90,30)(0,50)
\thicklines \put(-15,50){\line(1, 1){9}} \put( -4, 61){\line(1,
1){9}} \put(-15,70){\line(1,-1){20}}
\put(-20,60){\makebox(0,0)[cc]{$\Biggl\{$}}
\put(10,60){\makebox(0,0)[cc]{$\Biggr\}$}}
\put(18,60){\makebox(0,0){$=$}}
\put(30,61){\makebox(0,0){$-\frac12$}}
\put(40,60){\oval(20,20)[r]} \put(63,60){\oval(20,20)[l]}
\put(70,60){\makebox(0,0){$+\frac12$}}
\put(95,48){\oval(20,20)[t]} \put(95,72){\oval(20,20)[b]}
\put(-16,74){\makebox(0,0)[rb]{$G_1$}}
\put(6,74){\makebox(0,0)[lb]{$G_2$}}
\put(52,74){\makebox(0,0)[cb]{$G_I$}}
\put(95,74){\makebox(0,0)[cb]{$G_H$}}
\end{picture}
\label{Poisson-alg}
\ee
which was obtained in \cite{Gold} in the continuous parameterization.

A geodesic is said to be {\it graph simple\/} if it passes through each edge
of a graph no more than once. The set of graph simple geodesics is obviously
finite and depends on the choice of the combinatorial type of a
representing graph. These sets, although  obviously noninvariant under
the mapping class group action, form a convenient basis for closed
finite-dimensional geodesic algebras.

\begin{example}\label{exam4}{\rm
For the torus (${\cal T}^{}_{1,1}$), there are three generators $X,Y,Z$
satisfying the Poisson brackets
$$
\{X,Y\}=\{Y,Z\}=\{Z,X\}=2
$$
and with Casimir $X+Y+Z$. Correspondingly there are only
three graph simple geodesics whose geodesic functions are
\bea
\nonumber &{}&G_X=\tr
LX_YRX_Z=\e^{-Y/2-Z/2}+\e^{-Y/2+Z/2}+\e^{Y/2+Z/2},
\\
\label{torus-classic} &{}&G_Y=\tr
LX_ZRX_X=\e^{-Z/2-X/2}+\e^{-Z/2+X/2}+\e^{Z/2+X/2},
\\
\nonumber &{}&G_Z=\tr
LX_XRX_Y=\e^{-X/2-Y/2}+\e^{-X/2+Y/2}+\e^{X/2+Y/2}. \eea
Introducing the geodesic function $\wtd{G}_Z=\tr RX_ZRX_XLX_ZLX_Y$
obtained from $G_Z$ by the flip transformation (\ref{abc}),
it follows that
$\{G_X,G_Y\}={\wtd{G}_Z}/2-G_Z/2$, and since relation
(\ref{skeinclass}) implies that $G_XG_Y=G_Z+{\wtd G}_Z$, we
obtain
\be
\{G_X,G_Y\}=\frac12 G_XG_Y-G_Z,
\label{GXGYclass}
\ee
i.e., the classical Poisson algebra which closes on the subset
$\{G_X,G_Y,G_Z\}$ of geodesic functions (other relations are
obtained from cyclic permutations in (\ref{GXGYclass})) at the
price of introducing elements of the second order in the r.h.s.}
\end{example}

\newsection{Poisson algebras of geodesics}
In order to generalize Example~1 we must find a graph
on which graph simple geodesics constitute a convenient algebraic basis.
This is shown in Figure 2 where $m$ edges crosswise connect two horizontal line
subgraphs; note that the apparent vertices at the corners of these subgraphs
are not vertices. In this picture graph simple closed geodesics are those and
only those that pass along exactly two different ``vertical'' edges of the
graph; they are therefore enumerated by the numbers of these edges, and we
denote them by ${\cal G}_{ij}$ where $i<j$. The geodesic Poisson algebra
for ${\cal G}_{ij}$ is
\be
\{{\cal G}_{ij},{\cal G}_{kl}\}=\left\{\begin{array}{l}
0,\quad j<k,\\
0,\quad k<i,\ j<l,\\
{\cal G}_{ik}{\cal G}_{jl}-{\cal G}_{kj}{\cal G}_{il},\quad i<k<j<l,\\
\frac12 {\cal G}_{ij}{\cal G}_{jl}-{\cal G}_{il},\quad j=k,\\
{\cal G}_{jl}-\frac12 {\cal G}_{ij}{\cal G}_{il},\quad i=k,\ j<l\\
{\cal G}_{ik}-\frac12 {\cal G}_{ij}{\cal G}_{kj},\quad j=l,\ i<k.
\end{array}
\right. \label{P-geod} \ee
This algebra (\ref{P-geod}) is exactly that found by Nelson and Regge
\cite{NR1}.

The graph in Figure 2 has genus $\frac{m}{2}-1$ and {\it two\/} holes if $m$ is
even and genus $(m-1)/2$ and {\it one\/} hole if~$m$ is odd. Such geodesic
bases for $m$ even and {\it smooth\/} Riemann surfaces were considered in
\cite{NR}. The Poisson algebras of geodesics obtained there {\it coincide
exactly\/} with (\ref{P-geod}). These are the $so_q(m)$ algebras whose
representations were constructed in \cite{Klim}.

In the mathematical literature, the algebra (\ref{P-geod}) has also appeared as
the Poisson algebra of monodromy data (Stokes matrices) of some matrix
differential equation \cite{Ugaglia} and as the symplectic groupoid of
upper-triangular matrices~$A$ \cite{Bondal}. For $m\times m$-matrices, in general we
have $\left[\frac{m}{2}\right]$ central elements generated by the polynomial
invariants $f_{\cal G}(\lambda)\equiv \det({\cal G}+\lambda {\cal
G}^{T})=\sum_{}^{}f_i({\cal G})\lambda^i$. The total Poisson dimension $d$ of
the algebra (\ref{P-geod}) is $\frac{m(m-1)}{2}-\left[\frac{m}{2}\right]$, and
for $m=3,4,5,6,\dots$ we have $d=2,4,8,12,\dots\,$. The dimensions of the
corresponding Teichm\"uller spaces are $D=2,4,8,10,\dots\,$, so that the
Teichm\"uller spaces are embedded as the {\it Poisson leaves} into the algebra
(\ref{P-geod}). Starting from the genus 2 surface with two holes, the
dimensions of these special leaves become smaller than the highest dimensions
of the Poisson representations of (\ref{P-geod}).

$$
\setlength{\unitlength}{.4mm}%
\begin{picture}(90,120)(15,-15)
\thicklines
\put(0,95){\line(1, 0){125}}
\qbezier(0,95)(-25,95)(-10,85)
\qbezier(125,95)(150,95)(135,85)
\put(10,85){\line(1, 0){10}} \put(30,85){\line(1, 0){15}}
\put(55,85){\line(1, 0){15}} \put(80,85){\line(1, 0){15}}
\put(105,85){\line(1, 0){10}} \put(0,0){\line(1, 0){125}}
\qbezier(0,0)(-25,0)(-10,10)
\qbezier(125,0)(150,0)(135,10)
\put(10,10){\line(1, 0){10}} \put(30,10){\line(1, 0){15}}
\put(55,10){\line(1, 0){15}} \put(80,10){\line(1, 0){15}}
\put(105,10){\line(1, 0){10}} \put(-10,10){\line(5, 3){125}}
\put(10,10){\line(5, 3){125}} \put(20,10){\line(1, 1){15}}
\put(30,10){\line(1, 1){31}} \put(95,85){\line(-1, -1){31}}
\put(105,85){\line(-1, -1){15}} \put(45,10){\line(1, 3){7.5}}
\put(55,10){\line(1, 3){11.4}} \put(70,85){\line(-1, -3){11.4}}
\put(80,85){\line(-1, -3){7.5}} \put(-10,85){\line(5, -3){62.5}}
\put(10,85){\line(5, -3){25}} \put(115,10){\line(-5, 3){25}}
\put(135,10){\line(-5, 3){62.5}} \put(20,85){\line(1, -1){34.6}}
\put(30,85){\line(1, -1){21.5}} \put(95,10){\line(-1, 1){21.5}}
\put(105,10){\line(-1, 1){34.6}} \put(45,85){\line(1, -3){11.7}}
\put(55,85){\line(1, -3){7.5}} \put(70,10){\line(-1, 3){7.5}}
\put(80,10){\line(-1, 3){11.7}}
\thinlines
\qbezier(25,90)(20,90)(25,85)
\qbezier(100,90)(105,90)(100,85)
\qbezier(25,5)(20,5)(25,10)
\qbezier(100,5)(105,5)(100,10)
\put(25,90){\line(1,0){75}}
\put(25,5){\line(1,0){75}}
\put(25,85){\line(1,-1){22}}
\put(100,10){\line(-1,1){22}}
\put(100,85){\line(-1,-1){14}}
\put(25,10){\line(1,1){14}}
\put(10,45){\vector(1,-1){25}}
\put(115,45){\vector(-1,-1){25}}
\put(35,105){\makebox(0,0){$P^1_i$}}
\put(62.5,105){\makebox(0,0){$P^1_{i+1}$}}
\put(77.5,105){\makebox(0,0){$\cdots$}}
\put(92.5,105){\makebox(0,0){$P^1_{j-1}$}}
\put(35,-10){\makebox(0,0){$P^2_{j-1}$}}
\put(50.5,-10){\makebox(0,0){$\cdots$}}
\put(67.5,-10){\makebox(0,0){$P^2_{i+1}$}}
\put(92.5,-10){\makebox(0,0){$P^2_{i}$}}
\put(0,50){\makebox(0,0){$C_j$}}
\put(120,50){\makebox(0,0){$C_i$}}
\end{picture}
\label{octopus}
$$
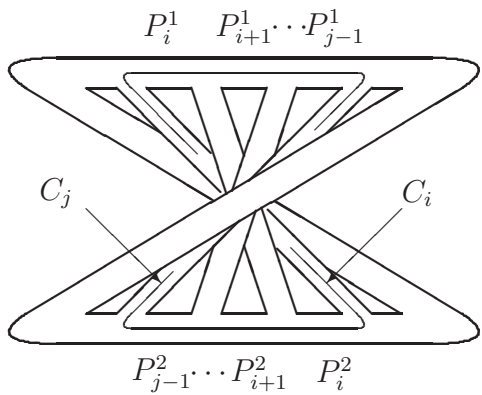
\centerline{\bf Figure 2}

\hspace{10pt}

\newsection{The free-field limit of Nelson--Regge algebras}
To produce an infinite-genus limit of geodesic algebras we consider graphs as
in Figure 2 with the number of ``rungs" of the crossed ladder tending to infinity
and replace the
index~$i$ by a continuous variable~$x$. With the parameterization (\ref{PZ}),
we introduce three fields $P^{1,2}(x)$ and $C(x)$ corresponding to the
respective quantities living on upper horizontal, lower horizontal, and
vertical edges. The form of the Poisson relations we wish to obtain can be
easily read off from equation (\ref{P-geod}). To be precise, let
$x_1<x_2<x_3<x_4$. Then,
\be
\label{pp1}
\{{\cal G}_{x_1,x_3},{\cal G}_{x_2,x_4}\}=
{\cal G}_{x_1,x_2}{\cal G}_{x_3,x_4}-{\cal G}_{x_1,x_4}{\cal
G}_{x_2,x_3},
\ee
where all quantities in the products on the r.h.s. mutually commute. In this
limit, we disregard all the boundary cases where some of the points $x_i$,
$i=1,\dots,4$, would coincide.

It is straightforward to calculate the geodesic function corresponding to the
path indicated in Figure 2 with (still discrete) variables $P^{1,2}_k$ and
$C_k$. We obtain
\bea
&{}&\tr LX_{P^2_i}R\cdots RX_{P^2_{j-2}}RX_{P^2_{j-1}}LX_{C_j}R
X_{P^1_{j-1}}L\cdots X_{P^1_{i+1}}LX_{P^1_{i}}RX_{C_i}
\nonumber
\\
&=&4s_1s_2\cosh (C_j/2)\cosh (C_i/2)+2\e^{C_i/2}\cosh (C_j/2)
\bigl(\e^{\Sigma^1+\Sigma^2}{+}s_1\e^{\Sigma^2}{+}s_2\e^{\Sigma^1}\bigr)
\nonumber
\\
&+&2\e^{-C_j/2}\cosh (C_i/2)
\bigl(\e^{-\Sigma^1-\Sigma^2}{+}s_1\e^{-\Sigma^2}{+}s_2\e^{-\Sigma^1}\bigr)
\nonumber
\\
&+&2\e^{C_i/2-C_j/2}\cosh(\Sigma^1-\Sigma^2),
\label{longsum}
\eea
where $\Sigma^{1,2}\equiv \sum_{k=i}^{j-1}P^{1,2}_k$ and
\be
\label{ss12}
s_1=\sum_{k=i+1}^{j-1}\e^{-\sum_{l=i}^{k-1}P^1_l/2+
\sum_{l=k}^{j-1}P^1_l/2},\quad
s_2=\sum_{k=i}^{j-2}\e^{-\sum_{l=i}^{k}P^2_l/2+
\sum_{l=k+1}^{j-1}P^2_l/2}.
\ee

In the limit of an infinite number of vertical lines, typical differences
$|i-j|$ become infinitely large, and in order to manage it
we introduce a scaling parameter
$\Delta$ that ensures that the combinations $\Delta|i-j|\equiv
|x_i-x_j|$ remain finite. To obtain continuously parameterized
expressions from (\ref{longsum}), we simultaneously rescale ${\cal
G}_{x_i,x_j}=\Delta^2\cdot {\cal G}_{ij}$ in order to replace summations
by integrations in (\ref{ss12}). Simultaneously,
we replace summations by integrations over a continuous variable
in the exponentials in the same formula.\footnote{Classically, this corresponds
to a situation where we also scale $P^{1,2}_i\to P^{1,2}_i/\Delta$, i.e., we
work in the domain of small variables $P_i$. However, we do not scale the Poisson (and
commutation) relations for these variables.}
The algebra (\ref{pp1}) is obviously invariant under such a rescaling,
since this reduces to the global additive renormalization of all
geodesic lengths $l(\gamma)$. In this field limit the expression
(\ref{longsum}) becomes drastically simplified, in fact only the first
term containing the product $s_1s_2$ survives, and we obtain
\bea
\nonumber
&{}&{\cal G}_{x_ix_j}=\int_{x_i}^{x_j}dy\int_{x_i}^{x_j}dz
\ 4\cosh (C(x_i)/2)\cosh (C(x_j)/2)\times
\\
&{}&\quad\times\exp\Bigl\{
\int_{x_i}^{x_j}d\xi\epsilon(y-\xi)P^1(\xi)d\xi/2
+\int_{x_i}^{x_j}d\rho\epsilon(z-\rho)P^2(\rho)d\rho/2\Bigr\},
\label{pp4}
\eea
where $\epsilon(x)=\{+1,x>0;\ -1,x<0\}$.

In this field limit, the field $C(x)$ can be scaled in two
possible ways. First, it becomes a single point when integrated over
the variable $x$. That is, if $C(x)$ remains quantum it cannot
contribute to the total expression (\ref{pp4}) since now the fields
$P^{1,2}(\xi)$ are macroscopic ensembles of the quantum fields
$P^{1,2}_k$. An alternative is to fix $C(x)$ to be a macroscopic
classical field. (This should be related to a deformation of the
initial Riemann surface in the continuous-genus limit; we will
discuss it elsewhere.) In this case, the fields $C(x)$ becomes {\it
commutative\/}. This also reflects the fact that in the continuous limit the
algebra (\ref{pp1}) becomes conformally invariant. The
only nontrivial commutation relations are now
\be
\label{pp5}
\{P^{1,2}(x),P^{1,2}(y)\}=\delta'(x-y),\qquad
\{P^1(x),P^2(y)\}\equiv 0,
\ee
i.e. the relations for string coordinates $\partial X(\sigma,\tau)$ at
coincident proper times $\tau$ (the free-field representation). That is, if
the variables $x,y$ lie in the interval $[0,2\pi]$, then by representing the
$P^{1,2}(x)$ as
$$
P^{1,2}(x)=\frac{1}{\sqrt{2\pi}}\sum_{l=-\infty}^{+\infty}\e^{ilx}\alpha^{1,2}_l,
$$
we find the standard CFT Poisson relations $\{\alpha^\mu_l,\alpha^\nu_p\}=l\delta_{l+p}\delta^{\mu\nu}$,
$\mu,\nu=1,2$.

It is convenient to introduce the integrated variables
$X^i(x)$ where $P^i(x)=\partial X^i(x)$. Their Poisson relations become
\be
\label{pp6}
\{X^{1,2}(x),X^{1,2}(y)\}=\epsilon(x-y)/2,\qquad
\{X^1(x),X^2(y)\}\equiv 0,
\ee
and then the quantities ${\cal G}_{x_ix_j}$ become integrals of local
field insertions:
\bea
\nonumber
{\cal G}_{x_ix_j}&=&\int_{x_i}^{x_j}dy\int_{x_i}^{x_j}dz\
4\cosh (C(x_i)/2)\cosh (C(x_j)/2)\times
\\
&{}&\times \e^{X^1(x_i)/2+X^1(x_j)/2+X^2(x_i)/2+X^2(x_j)/2-X^1(y)
-X^2(z)},
\label{pp7}
\eea
It can be checked that the Poisson relations (\ref{pp1}) follow from (\ref{pp7})
and (\ref{pp6}).

We now consider the boundary conditions for the free-field variables
$P^{1,2}(\xi)$. Consider a variation of Figure 2, where the
lower part of the picture is inverted i.e. twisted and opened out, so that all
rungs are twisted, and become vertical strips connecting the upper and
lower parts of the graph. The variables $P^{1,2}(\xi)$ live on these strips.
We briefly return to the discrete variables, $P^{1,2}_i$, $i=1,\dots,n$,
$n=m-3$, and let the variables on the two corners (twisted rungs) be
$P^1_{n+1}$ and $P^2_0$. It is easy to see that the only nonzero Poisson
brackets for the $P^{1,2}_i$ are
\be
\{P^1_i,P^1_{i+1}\}=\{P^2_{i-1},P^2_i\}=1,\quad i=1,\dots,n
\label{bc1}
\ee
and
\be
\{P^1_{n+1},P^2_n\}=\{P^1_{n+1},P^1_n\}=\{P^1_1,P^2_0\}=\{P^2_1,P^2_0\}=-1.
\label{bc2}
\ee

That is, the variables $P^1(\xi)$ and $P^2(\xi)$ exhibit a mirror-like
symmetry.

$$
\setlength{\unitlength}{.4mm}%
\begin{picture}(90,130)(10,-15)
\thicklines
\curve(0,50, -1,51, -2,53, -3,57, -3.5,60, -2.5,70, 3,80, 11,90, 25,96, 35,99, 50,100,
65,99, 75,96, 89,90, 97,80, 102.5,70, 103.5,60, 103,57, 102,53, 101,51, 100,50)
\curve(0,50, 1,49, 2,47, 3,44, 6,40, 12,30, 22,20, 30,15, 40,12, 50,10,
60,12, 70,15, 78,20, 88,30, 94,40, 97,44, 98,47, 99,49, 100,50)
\curve(1,51, 2,53, 3,56, 6,60, 12,70, 22,80, 30,85, 40,88, 50,90,
60,88, 70,85, 78,80, 88,70, 94,60, 97,56, 98,53, 99,51)
\curve(-1,49, -2,47, -3,43, -3.5,40, -2.5,30, 3,20, 11,10, 25,4, 35,1, 50,0,
65,1, 75,4, 89,10, 97,20, 102.5,30, 103.5,40, 103,43, 102,47, 101,49)
\thinlines
{\curvedashes[1mm]{2,1}
\curve(0,48, 0.5,43, 1,40, 5,30, 12,20, 25,10, 35,7, 50,5,
65,7, 75,10, 88,20, 95,30, 99,40, 99.5,43, 100,48)
}
\curve(15,22, 30,11, 50,7)
\curve(15,78, 30,89, 50,93)
\curve(15,78, 16,50, 15,22)
\curve(50,7, 50,50, 50,93)
\put(-5,85){\makebox(0,0){$P^1_1$}}
\put(50,110){\makebox(0,0){$P^1(\xi)$}}
\put(105,85){\makebox(0,0){$P^1_n$}}
\put(-5,15){\makebox(0,0){$P^2_1$}}
\put(50,-10){\makebox(0,0){$P^2(\xi)$}}
\put(105,15){\makebox(0,0){$P^2_n$}}
\put(-10,50){\makebox(0,0){$P^2_0$}}
\put(85,50){\makebox(0,0){$P^1_{n+1}$}}
\end{picture}
%
%
$$

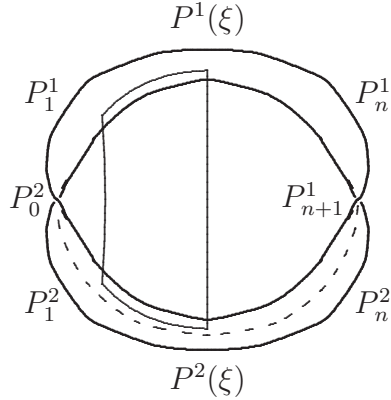
\centerline{\bf Figure 3}

\vspace{10pt}

For simplicity,
we discard in Figure~3 all the
$C$-variable rungs of the diagram (all of them are vertical twisted lines
joining symmetrical points $\xi$ on the upper and lower arcs on which the variables
$X^1(\xi)$ and $X^2(\xi)$ reside).
We keep only two such lines
with variables $P^2_0$ and $P^1_{n+1}$ in order to exhibit the possible
boundary conditions. It seems plausible that the boundary relations (\ref{bc2})
imply Neumann type boundary conditions for the string-like variables
$X^{1,2}(\sigma)$:
$$
P^1(0)\equiv\partial X^1(0)=P^2(0)\equiv \partial X^2(0)
$$
and
$$
P^1(2\pi)\equiv\partial X^1(2\pi)=P^2(2\pi)\equiv \partial X^2(2\pi).
$$

In the next section, we consider the quantization of the functions (\ref{pp7})
and construct their quantum algebra.

\newsection{Quantization}
Let ${\cal T}^\hbar_{g,n}({\Gamma})$, where ${\Gamma} \in {\Gamma}(S)$, be
the algebra generated by the generators $Z_\alpha^\hbar$ (one
generator per one (unoriented) edge $\alpha$) with the relations
\be
\label{comm}
[Z^\hbar_\alpha, Z^\hbar_\beta ] = 2\pi
\I\hbar\{Z_\alpha, Z_\beta\}
\ee
(cf.\ (\ref{WP-PB})) with the
$*$-structure $(Z^\hbar_\alpha)^*=Z^\hbar_\alpha$. Here
$Z_\alpha$  and $\{\cdot,\cdot\}$ stand for the respective
coordinate function and the Poisson bracket on the classical
Teichm\"uller space.

To any flip morphism, we now associate a homomorphism (Figure 1) of the
corresponding $*$-algebras that is idempotent and satisfies the five-term
relation \cite{ChF}. This morphism is given by (\ref{abc}) with the
(quantum) function \cite{Faddeev, kashaev}
\begin{equation} \label{phi}
\phi(z)\equiv \phi^\hbar(z) = -\frac{\pi\hbar}{2}\int_{\Omega}
\frac{\e^{-\I pz}}{\sinh(\pi p)\sinh(\pi \hbar p)}\D p,
\end{equation}
where the contour $\Omega$ goes along the real axis bypassing the
origin from above.

For any $\gamma$, the {\it quantum geodesic} operator $G^\hbar_\gamma$ is
\begin{equation} \label{qlen}
G^\hbar_\gamma \equiv \ORD{\tr P_{Z_1\dots Z_n}}\equiv
\sum_{{j\in J\atop \kappa\in\{j\}}} \exp\left\{{\frac{1}{2}
\sum_{\alpha \in E({\Gamma})} \bigl(m_j(\gamma,\alpha)
Z^\hbar_\alpha +2\pi \I\hbar c_j^\kappa(\gamma,\alpha)
\bigr)}\right\},
\end{equation}
where the {\it quantum ordering} $\ORD{\cdot}$ implies that we must change the
classical expression by introducing additional integer coefficients
$c_j^\kappa(\gamma,\alpha)$ determined from the following conditions.

{\bf1.} {\it The mapping class group} action ${\Delta}(S)$
(\ref{abc}) with $\phi^\hbar(z)$ from (\ref{phi})
preserves the set $\{G^\hbar_\gamma\}$, i.e., for any
$\delta \in {\Delta}(S)$ and any closed path $\gamma$, we have
$\delta(G^\hbar_\gamma) = G^\hbar_{\delta \gamma}$.

{\bf2.} {\it Geodesic algebra}. The product of two quantum
geodesics is a linear combination of {\it quantum geodesic
laminations} (QGLs) governed by the quantum skein relation \cite{Turaev}
below. In analogy with the classical case, a QGL is a set of self- and
mutually nonintersecting quantum geodesics.

{\bf3.} {\it Unorientedness.} Quantum traces of direct and inverse
geodesic operators coincide.

{\bf4.} If the closed paths $\gamma$ and $\gamma^\prime$ do not
intersect, then the operators $G^\hbar_\gamma$ and
$G^\hbar_{\gamma^\prime}$ commute. Therefore, the order of
quantum geodesic functions entering a QGL is irrelevant.

We introduce the Weyl ordering
${:}\e^{a_1}\e^{a_2}\cdots\e^{a_n}{:}\equiv \e^{a_1+\cdots+a_n}$ for any
set $\{a_i:\ a_i\ne-a_j\}$. It was found in \cite{Czech} that {\it for a
graph simple geodesic, the quantum ordering is the Weyl ordering\/}.

If $G^\hbar_1$ and $G^\hbar_2$ are two non selfintersecting quantum
geodesics with one nontrivial intersection, then
\be
\label{AG4}
G^\hbar_1G^\hbar_2
=\e^{-\I\pi\hbar/2}G^\hbar_Z+\e^{\I\pi\hbar/2}\wtd G^\hbar_Z.
\ee
That is, graphically, we obtain the Turaev \cite{Turaev} quantum skein
relation,
\be
\setlength{\unitlength}{.6mm}%
\begin{picture}(90,30)(0,50)
\thicklines
\put(-10,50){\line(1, 1){9}}
\put( 1, 61){\line(1,1){9}}
\put(-10,70){\line(1,-1){20}}
\put(16,60){\makebox(0,0){$=$}}
\put(30,61){\makebox(0,0){$\e^{-\I\pi\hbar/2}$}}
\put(40,60){\oval(20,20)[r]} \put(63,60){\oval(20,20)[l]}
\put(68,60){\makebox(0,0){$+$}}
\put(78,61){\makebox(0,0){$\e^{\I\pi\hbar/2}$}}
\put(95,48){\oval(20,20)[t]} \put(95,72){\oval(20,20)[b]}
\put(-11,74){\makebox(0,0)[rb]{$G^\hbar_1$}}
\put(11,74){\makebox(0,0)[lb]{$G^\hbar_2$}}
\put(52,74){\makebox(0,0)[cb]{$G^\hbar_Z$}}
\put(95,74){\makebox(0,0)[cb]{$\wtd G^\hbar_Z$}}
\end{picture}
\label{skein}
\ee
In (\ref{skein}) the order of the crossing of $G^\hbar_1$ and $G^\hbar_2$
indicates which geodesic occupies the first place in the product; the
rest of the graph is unchanged).

If two QGLs have {\it more\/} than one intersection, the quantum skein
relations must be applied {\it simultaneously\/} at {\it all\/}
intersection points \cite{Czech}. In particular,
this implies the Riedemeister moves for a graph, if we set
the empty loop to be $-\e^{-i\pi\hbar}-\e^{i\pi\hbar}$.

The algebra (\ref{P-geod}) was quantized by the deformation
quantization method in \cite{NR,NRZ}. By imposing the quantization
conditions (\ref{comm}), we obtain the quantum $so_q(m)$ algebra
of \cite{NRZ}: $a_{ij}$ commutes with $a_{kl}$ if $i<j<k<l$ or
$i<k<l<j$; the elements $a_{ij}$, $a_{jk}$, and $a_{ik}$
constitute the $so_q(3)$ subalgebra; and if
$i<k<j<l$, then $[a_{ij},a_{kl}]=\xi(a_{ik}a_{jl}-a_{il}a_{jk})$, where
$\xi=\e^{-i \pi\hbar}-\e^{i \pi\hbar}$.

The {\it quantum\/} operators ${\cal G}^\hbar_{x_i,x_j}$ are determined by
equation (\ref{pp7}) when the Weyl ordering is assumed, and the
commutation relations between the fields appearing there are
\bea
\label{ss21}
&{}&[X^{1,2}(x),X^{1,2}(y)]=\pi\I\hbar\epsilon(x-y),
\\
\nonumber
&{}&[X^1(x),X^2(y)]=[X^{1,2}(x),C(y)]=[C(x),C(y)]=0.
\eea
It is straightforward to check that (again assuming
$x_1<x_2<x_3<x_4$)
\be
\label{ss22}
\bigl[{\cal G}^\hbar_{x_1,x_3},{\cal G}^\hbar_{x_2,x_4}\bigr]=
\xi \bigl({\cal G}^\hbar_{x_1,x_2}{\cal G}^\hbar_{x_3,x_4}
-{\cal G}^\hbar_{x_1,x_4}{\cal G}^\hbar_{x_2,x_3}\bigr),
\ee
where, as previously, the quantities in the products in the r.h.s. commute.

\newsection{Conclusions}  This paper should be considered as a first step in
exploring the possibilities of using infinite/continuous genus limits of
Riemann surfaces in 3D gravity to describe the effects of CFT/string theory.
For example, one can interpret the continuous-genus limit classically as
describing surfaces similar to two sphere-like regions (the variables
$P^{1,2}(\xi)$) connected by infinitely many handles (the variables $C(\xi)$).
It would be interesting to find an analogous treatment of the more complicated
algebras of $SL(n,{\Bbb R})$ type (see~\cite{FG}) for which similar
constructions may have some meaning.

An interesting question is to find eigenfunctions
for the quantum geodesic operators. The spectrum is
obviously continuous and positive definite, and finding possible coherent states for operators (\ref{pp7})
in terms of string variables seems to be interesting problem with possible applications
in conformal field theory. We expect
that the fermionization technique can be useful. This work is in progress.
Also, worth noting is the
{\em doubling} of variables (we have two sets, $X^1(\xi)$ and $X^2(\xi)$ of them). Note that
the algebra (\ref{ss22}) holds only when we have one or two sets of variables; adding
extra (string) variables destroys this structure.

\bigskip

\noindent {\bf Acknowledgements}

\noindent {\small ~This work was supported in part by RFBR Grant
No.~05-01-00498, Support Grant for Scientific Schools NSh-2052.2003.1,
Programme Mathematical Methods of Nonlinear Dynamics, Istituto Nazionale di
Fisica Nucleare (INFN) of Italy Iniziativa Specifica FI41, and the Italian
Ministero dell' Universit\`a e della  Ricerca Scientifica e Tecnologica (MIUR)
under contract PRIN-2003023852 "Physics of Fundamental Interactions: gauge
theories, gravity and strings".}


\bigskip

\begin{thebibliography}{99}
\bibitem{VV}
Verlinde E.P. and Verlinde H.L., ``Conformal Field Theory and Geometric
Quantization'', in {\it Superstrings '89} (Trieste 1989) World Scientific
Publishing, River Edge, NJ, 1990, pp.422-449.
\bibitem{ChF}
Chekhov L.O. and Fock V.V., ``Quantum Teichm\"uller Space'', Theor. Math. Phys.
{\bf 120} (1999) 1245-1259.
\bibitem{Ch-Fock-TMF}
Chekhov L.O. and  Fock V.V., ``Quantum Mapping Class Group, Pentagon Relation, and
Geodesics'', Proc. Steklov Math. Inst. {\bf 226} (1999) 149-163.
\bibitem{Penner}
Penner R.C., ``The decorated Teichm\"uller space of punctured surfaces",
Commun. Math. Phys. {\bf 113} (1987) 299-339.
\bibitem{Fock1}
Fock V.V., ``Combinatorial description of the moduli space of
projective structures'', hepth/9312193.
\bibitem{Gold}
Goldman W.M.: ``Invariant functions on Lie groups and Hamiltonian flows of
surface group representation'', Invent. Math. {\bf85} (1986) 263-302.
\bibitem{NR}
Nelson J.E. and Regge T., ``Homotopy Groups and $2+1$ dimensional Quantum Gravity'',
Nucl. Phys. {\bf B328} (1989) 190-202; Nelson J.E., Regge T. and Zertuche F.,
``Homotopy Groups and $2+1$ dimensional Quantum De Sitter Gravity,'', Nucl.
Phys. {\bf B339} (1990) 516-532.
\bibitem{NRZ}
Nelson J.E. and Regge T., ``$2+1$ Quantum Gravity'', Phys. Lett. {\bf B272} (1991)
213-216; ``Invariants of 2+1 Gravity'', Commun. Math. Phys. {\bf 155} (1993)
561-568.
\bibitem{NR1}
Nelson J.E. and Regge T., ``$2+1$ Quantum Gravity for High Genus'', Class. Qu.
Grav. {\bf 9} (1992) S187-S196.
\bibitem{Klim}
Havl\'i\v cek M., Klimyk A.V. and Po\v sta S., ``Representations of
the cyclically symmetric q-deformed algebra so[sub q](3)'', J. Math. Phys.
{\bf 40} (1999) 2135;~Fairlie D.B., `` Quantum Deformations of SU(2)'',
J.Phys. {\bf A23} (1990) L183.
\bibitem{Ugaglia}
Ugaglia M.,``On a Poisson structure on the space of Stokes matrices'',
Int. Math. Res. Not. {\bf1999}, No.~9, (1999) 473.
\bibitem{Bondal}
Bondal A., ``A symplectic groupoid of
triangular bilinear forms and the braid groups'', preprint IHES/M/00/02.
\bibitem{Faddeev}
Faddeev L.D., ``Discrete Heisenberg-Weyl Group and Modular Group'', Lett.
Math. Phys. {\bf34} (1995) 249-254.
\bibitem{kashaev}
Kashaev R.M., ``Quantization of Teichm\"uller Spaces and the Quantum
Dilogarithm'', Lett. Math. Phys. {\bf43} (1998) 105-115.
\bibitem{Czech}
Chekhov L.O. and Fock V.V., ``Observables in 3D Gravity and Geodesic Algebras'',
Czechoslovak J. Phys. {\bf50} (2000) 1201-1208.
\bibitem{Turaev}
Turaev V.G., ``Skein quantization of Poisson algebras of loops on surfaces'',
Ann. Sci. \'Ec. Norm. Sup., Ser.~4 {\bf24} (1991) 635-704.
\bibitem{FG}
Fock V.V. and Goncharov A.B., ``Cluster ensembles, quantization and the
dilogarithm'', math.AG/0311245.
\end{thebibliography}
\end{document}